\documentclass[aps,prx,reprint,superscriptaddress]{revtex4-2}

\usepackage{graphicx}  

\usepackage{longtable}
\usepackage[T1]{fontenc}
\usepackage{dcolumn}   
\usepackage[inline]{enumitem}

\usepackage{bm}        
\usepackage{amsfonts}  
\usepackage{amsmath}   
\usepackage{amssymb}   

\usepackage{amsthm}
\usepackage{tikz}
\usetikzlibrary{quantikz}
\usepackage{glossaries}

\usepackage{physics}

\usepackage{color}
\usepackage[colorlinks=true,allcolors=blue]{hyperref}

\newcommand{\pwisein}{\left\{ \begin{array}{ll}}
\newcommand{\pwiseout}{\end{array}\right.}

\begin{document}

\title{Optimization of current-based sensing of phonon temperature using a finite reservoir}
\author{Sindre Brattegard} 
\email{brattegs@tcd.ie}
\affiliation{School of Physics, Trinity College Dublin, College Green, Dublin 2, Ireland}

\author{Stephanie Matern}
\email{materns01@gmail.com}
\affiliation{Pitaevskii BEC Center, CNR-INO and Dipartimento di Fisica, Università di Trento, 38123 Trento, Italy}

\author{Mark T. Mitchison}
\email{mark.mitchison@kcl.ac.uk}
\affiliation{Department of Physics, King’s College London, Strand, London, WC2R 2LS, United Kingdom}
\affiliation{School of Physics, Trinity College Dublin, College Green, Dublin 2, Ireland}

\author{Saulo V. Moreira} 
\email{moreirsv@tcd.ie}
\affiliation{School of Physics, Trinity College Dublin, College Green, Dublin 2, Ireland}

\begin{abstract}
    
    In realistic nanoscale transport set-ups, electron-phonon coupling leads to the exchange of heat between phonon baths and electronic reservoirs with finite heat capacities. Such exchange affects the finite reservoir's temperature.
    However, this sensitivity of the finite reservoir temperature to the exchange of heat with the finite reservoir has remained unexplored for thermometry. Here, we fill this gap by combining current metrology techniques with a thermodynamic framework encompassing finite reservoirs. We focus on an experimentally realizable set-up with a quantum dot coupled to a finite reservoir and consider two distinct current-based strategies in the long time limit, namely monitoring quanta exchanged between the quantum dot and finite reservoir and the measurement of the total current flowing from the quantum dot into an infinite reservoir. A third strategy involves measurements of the quantum dot occupation. For a large but finite reservoir, we show that the Fisher information for all three strategies captures the finite reservoir's contribution to sensitivity through common factors. We also demonstrate that monitoring quanta exchanged between the system and finite reservoir in the long time limit achieves optimal precision. Finally, we provide an optimization analysis that explores how maximal precision can be achieved within each of the current-based strategies by tuning the gate voltage.

\end{abstract}


\maketitle 


\textit{Introduction}.--- Measuring the temperature of phonons is fundamental for understanding thermoelectric properties influencing the performance of quantum electronic devices. Devising practical high-precision phonon temperature measurements is, nonetheless, challenging.
In experiments, the phonon temperature is often inferred using spectroscopic methods, in particular Raman spectroscopy~\cite{Beechem2007, Beechem2015}. While commonplace, Raman thermometry techniques have relatively low precision and can be sensitive to imperfections in the sample~\cite{Sandell2020}.

In nonequilibrium experimental configurations, heat exchanges between electron and phonon reservoirs enabled by electron-phonon coupling carry information about the phonon bath's parameters. Focusing on measurements of transport observables, such as currents, could, therefore, potentially lead to advantageous estimation schemes. Most importantly, current observables are experimentally accessible in a variety of open-system platforms at the nanoscale. 

In this vein, current-measurement-based parameter estimation strategies recently started to gain traction. Examples include measuring currents for the estimation of system or environment parameters in nonequilibrium scenarios, such as thermometry exploiting quantum thermal machines~\cite{Hofer2017} and quantum dot (QD) set-ups~\cite{Maradan2014, Yang2019}. Further recent theoretical advances include the derivation of classical and quantum parameter estimation bounds in mesoscopic electronic settings employing an adiabatic linear-response approach~\cite{mihailescu2025} and the Landauer-B\"uttiker formalism~\cite{Khandelwal2025}. As well as this, Bayesian approaches for continuously measured two-level systems coupled to fermionic or bosonic environments have been investigated~\cite{Boeyens2023}.

Previous works on current-based metrology have approximated the environments driving the system out of equilibrium as being infinitely large. Notably, however, many realistic experimental transport setups encompass finite-size reservoirs~\cite{Pekola2021, Dutta2020, Spiecker2023, Spiecker2024, Champain2024}, which undergo fast internal thermalization due to electron-electron interactions~\cite{Pothier1997, Schmidt2004}. Due to its finite heat capacity, the temperature of such a finite reservoir is affected by exchanges of heat with the surroundings. This feature has been shown to influence transport properties~\cite{Gallego2014, Schaller2014, Grenier2016,Amato2020,Matern2024}, nonequilibrium thermodynamics~\cite{Strasberg2021-1, RieraCampeny2021,RieraCampeny2022, Elouard2022,Moreira2023}, and heat-engine performance~\cite{Yuan2026, Strasberg2021-2,Mamede2026}, but it also makes finite reservoirs ideal platforms for calorimetry~\cite{Donvil2022}. Indeed, recent experiments have demonstrated electronic nano-calorimeters that can detect tiny amounts of heat exchange via measurable changes in their temperature~\cite{Brange2018, Karimi2020}. This sensitive relationship between heat exchange and temperature naturally suggests that finite reservoirs could also be exploited for current-based thermometry, yet to date this possibility has remained unexplored in the literature.

In this letter, we investigate phonon temperature estimation by combining current-based metrology techniques with transport theory encompassing finite-size reservoirs. We focus on an experimentally realizable non-equilibrium transport set-up supporting particle and heat currents, where an electronic finite reservoir is both coupled to a bath of phonons and a single-level QD.
Our analysis considers two distinct current-measurement-based strategies in the long time limit: (\emph{I})~monitoring quanta exchanged between the QD and the baths, and (\emph{II}) measuring the net particle current going through the QD into an infinite reservoir. 
Pivoting from current measurements, we consider a third strategy, i.e., (\emph{III}) measuring the QD occupation.
We derive analytical expressions for the Fisher information for these three strategies under the assumption of a large, but still finite, reservoir. All three formulas are shown to share factors revealing how finite reservoirs can be harnessed for thermometry. 
Furthermore, while we show that strategy \emph{I} enables the optimal performance achievable in our set-up, our optimization analysis further explores how maximal precision can be achieved by tuning the gate voltage within each current-based strategy (\emph{I} and \emph{II}).

\begin{figure}
    \centering
    \includegraphics[width=0.9\linewidth]{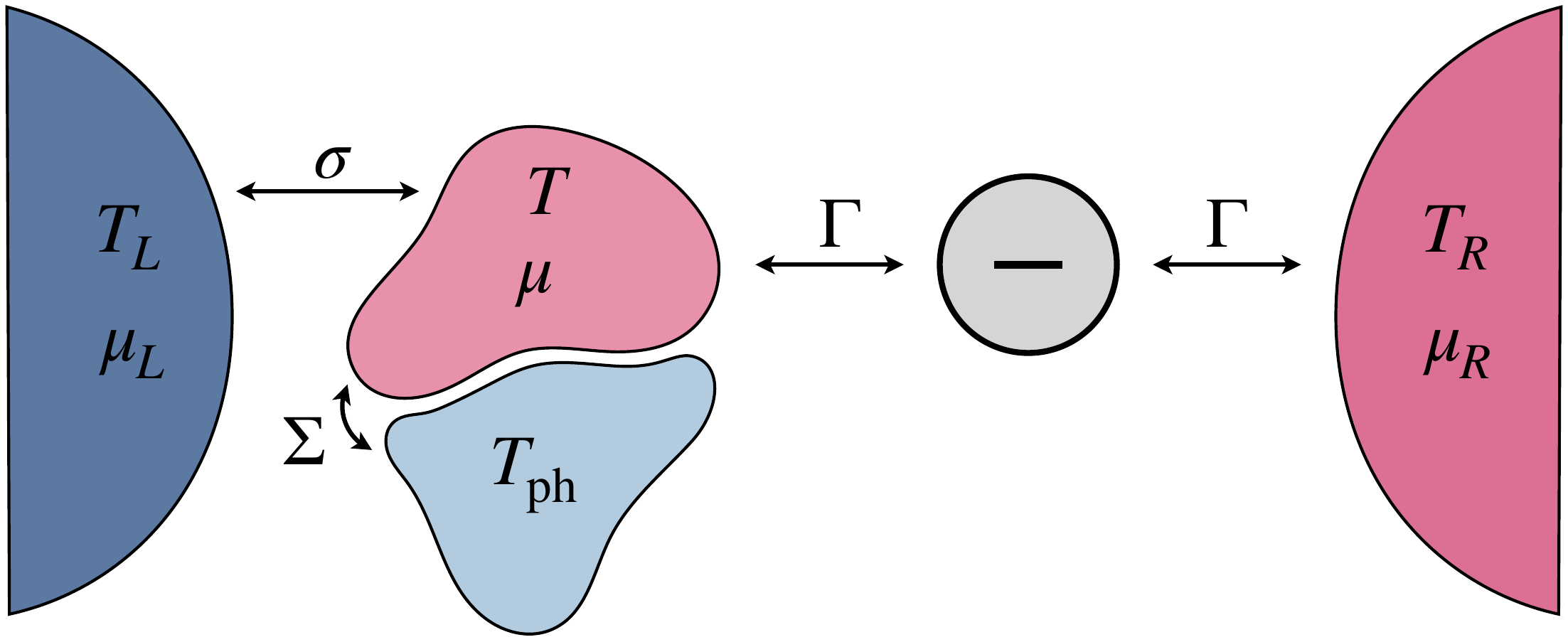}
    \caption{Sketch of the finite-size reservoir at stationary temperature and chemical potential, $T$ and $\mu$, coupled to a phonon bath at temperature $T_{\text{ph}}$ and to the quantum dot system. The finite reservoir is also coupled to the left infinite-size reservoir at temperature $T_L$ and $\mu_L$, while the quantum dot is coupled to another infinite-size reservoir to its right, at temperature $T_R$ and chemical potential $\mu_R$.}     \label{fig:sketch}
\end{figure}

\textit{Nonequilibrium transport set-up}.--- We consider the set-up sketched in Fig.~\ref{fig:sketch}, where the electronic finite reservoir-QD is coupled to left (L) and right (R) infinite-size reservoirs at different temperatures $T_{L/R}$ and chemical potentials $\mu_{L/R}$. Furthermore, the finite reservoir exchanges heat with a bath of phonons at some temperature $T_\mathrm{ph}$ as depicted in Fig.~\ref{fig:sketch}.
We take all electronic reservoirs to consist of non-interacting electrons, described by Fermi-Dirac distributions $f(E, \tilde{T}, \tilde{\mu}) = \left[1+e^{(E-\tilde{\mu})/\tilde{T}}\right]^{-1}$ (we set $k_B=1$). 
The finite reservoir's thermalization via electron-electron interaction is considered to be very fast, so that it is in a quasiequilibrium state at all times, with well-defined, time-dependent temperature $T(t)$ and chemical potential $\mu(t)$,
characterized by $f(E, T(t), \mu(t))$~\cite{Berg2015, Karimi2020}.

The time evolution of the system is described by the following phenomenological master equation~\cite{Ingold1992}
\begin{align}\label{RateEq}
    \frac{d}{dt}\begin{bmatrix} p_0 \\ p_1 \end{bmatrix} = \mathcal{W} \begin{bmatrix} p_0 \\ p_1 \end{bmatrix} =  \sum_{\alpha =  L,R}
    \begin{bmatrix}
        -\Gamma_\alpha^{\text{in}} & \Gamma^\text{out}_\alpha  \\
        \Gamma^\text{in}_\alpha &  - \Gamma^\text{out}_\alpha 
    \end{bmatrix}
    \begin{bmatrix} p_0 \\ p_1 \end{bmatrix},
\end{align}
where $p_{n}$ denotes the probability of the QD being in state $n=\{0,1\}$, i.e., empty ($n=0$) or occupied ($n=1$). The tunnelling rates for electrons to jump in or out of the QD are given by~\cite{Golovach2011}
\begin{align}
    &\Gamma^\text{in}_L = \Gamma f(\epsilon, T, \mu), \qquad  \Gamma^\text{out}_L = \Gamma[1 - f(\epsilon, T, \mu)] \\
&\Gamma^\text{in}_R = \Gamma f(\epsilon, T_R, \mu_R), \quad 
\Gamma^\text{out}_R = \Gamma[1 - f(\epsilon, T_R, \mu_R)].
\end{align} 
The master equation in Eq.~\eqref{RateEq} is valid in the weak coupling regime when $\Gamma \ll T$.
Average particle and heat currents, denoted by $\langle I_\mathrm{QD}^{(L/R)}\rangle$ and $\langle\dot{Q}_\mathrm{QD}^{(L/R)}\rangle$, respectively, flowing from the QD into the finite reservoir (L) and infinite reservoir (R), can be expressed as
\begin{align}
    \langle I_\mathrm{QD}^{(L/R)}\rangle &= p_1\Gamma_{L/R}^\mathrm{out} - p_0\Gamma_{L/R}^\mathrm{in} \\
\langle\dot{Q}_\mathrm{QD}^{(L/R)}\rangle &= (\epsilon - \mu)\langle I_\mathrm{QD}^{(L/R)} \rangle. \label{Q_QD}
\end{align}
Furthermore, we consider linear response equations to describe transport between the infinite and finite reservoirs~\cite{Matern2024}. The corresponding particle and heat currents are given by
\begin{align}
\label{eq:inf_currents}
    \langle I_\mathrm{inf}\rangle = \sigma\Delta\mu, \quad
    \langle \dot{Q}_\mathrm{inf}\rangle = \frac{\pi^2\Bar{T}\sigma}{3}\Delta T +\Bar{\mu} \sigma \Delta \mu,
\end{align}
where $\Bar{\mu} = (\mu_L + \mu)/2$ and $\Bar{T} = (T_L+T)/2$ are the average chemical potential and temperature, and  $\Delta \mu = \mu_L - \mu$ and $\Delta T = T_L - T$ are chemical potential and temperature differences.
These expressions assume energy-independent transmission between the finite and infinite electron reservoirs, and are valid when the electrical conductance $\sigma$ is much larger than  $\Delta \mu $ and $\Delta T$. In the Supplemental Information (SI)~\cite{SI}, we provide a detailed analysis of their regime of validity using the Landauer-Büttiker formalism.

Finally, the exchange of heat between the finite electronic reservoir and the phonon bath is expressed as
\begin{align}\label{Q_ph}
    \langle\dot{Q}\rangle = -\Sigma(T^n - T_\mathrm{ph}^n),
\end{align}
where $T_\mathrm{ph}$ is the temperature of the phonon bath and $\Sigma$ accounts for the phonon-electron interaction. Eq.~\eqref{Q_ph} is valid when we only consider a single phonon mode with a well-defined temperature $T_\mathrm{ph}$, and is valid to first order in $\Sigma$.
Here, we only consider a single phonon mode with a well-defined temperature $T_{\mathrm{ph}}$. The extension to multiple phonon modes is, however, straightforward within our framework.
The exponent $n$ takes different values depending on the system. Usually, $n=5$ for normal bulk metals~\cite{pekola_quantum_2018, Brange2018}, but other values are possible for dimensionally restricted electron gases  \cite{Chow1996, Schmidt2004, Giazotto2006,wiesner_electronphonon_2022,singh_detailed_2013}.

From here on, we write $\langle I_{\mathrm{QD}}\rangle = \langle I_{\mathrm{QD}}^L\rangle$ and $\langle \dot{Q}_{\mathrm{QD}}\rangle = \langle \dot{Q}_{\mathrm{QD}}^L\rangle$ to ease the notation.
The time evolution of the  finite reservoir's chemical potential and temperature will depend on particle and heat currents as~\cite{Matern2024}
\begin{align}\label{MuDot}
    & \dot{\mu} = \frac{1}{\nu_0}\left(\expval{I_\mathrm{QD}} + \expval{I_\mathrm{inf}} \right), \\
    \label{TDot}
    & \dot{T} = \frac{1}{C_\mathrm{el}} (\expval*{\dot{Q}_\mathrm{QD}} + \expval*{\dot{Q}_\mathrm{inf}} + \expval*{\dot{Q}_\mathrm{ph}} ),
\end{align}
where $\nu_0$ is the density of states of the electrons (taken to be energy independent for simplicity), and~$C_\mathrm{el} = \nu_0 \pi^2T/3$ is the specific heat of a free electron gas. 
At the nonequilibrium stationary state (NESS), we have that $\dot{\mu} = \dot{T} = 0$. From now on, $T$ and $\mu$ will denote fixed, \textit{stationary} temperature and chemical potential of the finite reservoir, which can be found by self-consistently solving the coupled equations for the currents presented above.

The finite reservoir's NESS temperature and chemical potential depend on how strongly the finite reservoir couples to the infinite reservoir, the QD, and the phonon reservoir. In the following, we define the dimensionless quantity $\zeta = \sigma T_L/\Gamma$, representing the coupling strength of the finite reservoir to the infinite electron reservoir relative to its coupling to the QD. We also define $\xi = \Sigma T_\mathrm{ph}^n/\sigma T_L^2$, which captures the relative coupling strength of the finite reservoir to the phonon versus the electron reservoir. To make progress analytically, we assume that $\zeta \gg 1$ and $\xi\gg 1$, which describes the regime of a large, but still finite, reservoir~\cite{Matern2024} whose temperature is significantly influenced by the phonon bath. With these approximations, we can write $T = T_{\mathrm{ph}} + \Delta T$ and $\mu = \mu_L + \Delta \mu$, with $\Delta T/T_{\mathrm{ph}}\ll 1$ and $\Delta \mu/\mu_L\ll 1$ given by
\begin{align}
    \Delta\mu \approx -\frac{V_b}{\sigma}G, \qquad
    \Delta T \approx \frac{\pi^2T_{\mathrm{ph}}}{6n\xi}\left(1-\frac{T_{\mathrm{ph}}^2}{T_L^2}\right)\label{deltas},
\end{align}
where $G$ is the conductivity of transport through the QD. See the SI~\cite{SI} for the detailed derivation and justification of the expressions in Eq.~\eqref{deltas}. The SI also provides an analysis of the contrasting regime in which $\xi \ll 1$ (ultra-weak coupling to the phonon reservoir), but this yields substantially reduced thermometric sensitivity so we focus on the case of $\xi \gg 1$ from here on. 

A few remarks about~\eqref{deltas} are in order. We first note that $\Delta \mu$ is insensitive to the phonon temperature. This results from the fact that the phonon temperature enters~\eqref{MuDot} indirectly, through $T$.
Instead, the change in $\mu$ is dictated by the relation between the tunneling strength into the QD and the infinite reservoir. 
On the other hand, $\Delta T$ is inversely proportional to
the relative strength $\xi$.

\textit{Estimation theory}.--- Given a sequence of measurement outcomes, $\mathbf{x} = \{x_1,x_2, \dots x_N\}$, an estimate of the phonon temperature can be obtained by employing an estimator, which is a function of the measurement data $\mathbf{x}$.
In particular, \emph{unbiased} estimators, $\check{T}_\mathrm{ph}(\mathbf{x})$,
satisfy the property that their expected values, taken over all possible measurement records, correspond to the true value of the parameter, $\langle \check{T}_\mathrm{ph}(\mathbf{x}) 
\rangle = T_{\rm ph}$. The precision of the phonon temperature estimation is captured by the variance of $\check{T}_\mathrm{ph}(\mathbf{x})$, which is bounded by the Cramér–Rao bound~\cite{Cramer1946, Rao1973} as $\mathrm{Var}(\check{T}_{\mathrm{ph}}(\mathbf{x})) \geq 1/F(T_\mathrm{ph})$, where $F(T_\mathrm{ph})$ is the Fisher information~\cite{Kay2013}
\begin{equation}
    F(T_{\mathrm{ph}}) = \sum_{\mathbf{x}} P(\mathbf{x})\left[\pdv{}{T_{\mathrm{ph}}}\ln P(\mathbf{x})\right]^2.
\end{equation}
Here, $P(\mathbf{x})$ is the $T_{\mathrm{ph}}$-dependent probability to obtain the measurement data $\mathbf{x}$.

\textit{Strategy I}.--- We first consider continuously measuring the system, as it evolves under Eq.~\eqref{RateEq}. In this case, measurement outcomes are, in general, correlated with each other~\cite{Radaelli2023}. A measurement record is, in this way, constituted of data points characterizing a given \emph{trajectory}. For example, trajectories 
 can be constructed by keeping track of the sequence of stochastic jumps of electrons tunnel in and out of the QD, $\mathbf{x} = \{x_1, x_2, \dots, x_N \}$, during the time interval $0 \le t \le \tau$. 
 
In this case, for a sufficiently large $\tau$, the Fisher information can be expressed as~\cite{Smiga2023}
\begin{equation}\label{FIcontinuous}
    F_I(T_{\mathrm{ph}}) =   
\left[(\partial_{T_{\mathrm{ph}}}{\Gamma_L^{\mathrm{in}}})^2  \frac{p_0^{\mathrm{ss}}}{\Gamma_L^{\mathrm{in}}} + (\partial_{T_{\mathrm{ph}}}\Gamma_L^{\mathrm{out}})^2  \frac{p_1^{\mathrm{ss}}}{\Gamma_L^{\mathrm{out}}}\right] \tau,
\end{equation}
where $p_1^{\mathrm{ss}} = (f+f_R)/2$ and $p_0^{\mathrm{ss}} = 1 - p_1^{\mathrm{ss}}$ are stationary probabilities.
We note that $F_I$ depends only on the rates describing the coupling to the finite reservoir, $\Gamma_{L}^{\mathrm{in}}, \Gamma_{L}^{\mathrm{out}}$, as the derivatives of $\Gamma_{R}^{\mathrm{in}}, \Gamma_{R}^{\mathrm{out}}$ with respect to the phonon temperature are zero. In this way, the sensitivity is entirely determined by the exchange of quanta with the finite reservoir.
In the SI~\cite{SI}, by considering the approximations in~\eqref{deltas}, we show that $F_I$ can be approximated as
\begin{equation}\label{ApproxF_I}
    F_I (T_{\mathrm{ph}}) \approx \chi_e\chi_{\mathrm{ph}} \left[\frac{\langle N_{\mathrm{in}}\rangle}{f^2} + \frac{\langle N_{\mathrm{out}}\rangle}{\bar{f}^2}  \right],
\end{equation}
where $f = f(\epsilon, T, \mu)$ and $\bar{f} = 1 - f(\epsilon, T, \mu)$, while $\langle N_{\mathrm{in}} \rangle = p_0^{\mathrm{ss}}\Gamma_L^{\mathrm{in}}\tau$ and $\langle N_{\mathrm{out}} \rangle = p_1^{\mathrm{ss}}\Gamma_L^{\mathrm{out}}\tau$. Additionally, we have that 
\begin{align}
    \chi_e  \equiv& \left(\pdv{f} {T}\right)^2= \frac{(\epsilon -\mu)^2 }{16T^4\cosh^4(\frac{\epsilon - \mu}{2T})}, \label{chi_e} \\
   \chi_{\mathrm{ph}} \equiv& \left(\pdv{T} {T_{\mathrm{ph}}}\right)^2=\left[1 +\frac{\pi^2}{6n\xi}\left((1-n)+\frac{T_{\mathrm{ph}}^2}{T_L^2}(n-3)\right)\right]^2. \label{chi_ph}
\end{align}

The expression in Eq.~\eqref{ApproxF_I} is particularly insightful, as each of the three multiplicative factors captures distinct contributions to the phonon temperature sensitivity, as explained in the following. The first factor, $\chi_e$, quantifies the sensitivity of the finite reservoir's Fermi-Dirac occupation, $f(\epsilon, T, \mu)$, to changes in its temperature $T$. The second factor, $\chi_{\mathrm{ph}}$, measures how $T$ changes with the phonon temperature. Altogether, the product $\chi_e\chi_{\mathrm{ph}}$ weights the Fisher information with the sensitivity of the finite reservoir to the phonon temperature. In this sense, we show that the finite reservoir temperature becomes independent of $T_{\mathrm{ph}}$ when $\xi \rightarrow 0$ and $\zeta \rightarrow 0$. This means that
$\chi_{\mathrm{ph}}$ vanishes in the limit of an infinitely large electronic reservoir, which implies zero sensitivity (see the SI~\cite{SI}).
Finally, the last factor is expressed in terms of the expected value of the number of tunneling events into (out) the QD from (into) the finite reservoir during the measuring time $\tau$, $\langle N_{\mathrm{in}} \rangle$ ($\langle N_{\mathrm{out}} \rangle)$.

To put our result in Eq.~\eqref{FIcontinuous} into perspective, we consider the upper bound on the Fisher information, namely the \emph{quantum Fisher information}~\cite{PARIS_quantum_estimation}, derived by Gammelmark and Mølmer~\cite{Gammelmark2014} for measurements on the global system (i.e., system and environment) during a certain time $\tau$. Specifically, they obtained an analytical expression for the Fisher information corresponding to the optimal measurement on the global system, out of all possible quantum measurements. This result therefore corresponds to the maximum precision achievable by measuring the global system for a sufficiently long time (see the SI~\cite{SI} for details).

In the SI~\cite{SI}, we show that, in the long time limit, the Fisher information for jump trajectories equals the upper bound derived in Ref.~\cite{Gammelmark2014} for any classical rate equation. We can therefore conclude that, for very large $\tau$, monitoring jumps corresponds to the best possible precision.

\textit{Strategy II}.---
We will now consider measuring the net charge $N$ transferred from the QD into the right reservoir during the time $\tau$. Such a measurement can be performed by integrating the DC current measured by an ammeter over a time window $\tau$, and can be modeled theoretically as a two-point measurement of the number of electrons in a given reservoir~\cite{Esposito2009}. For large $\tau$, the average and fluctuations of $N$ increase linearly with time, i.e., $\langle N\rangle = \tau \langle I_{\mathrm{QD}}\rangle$ and $\mathrm{Var}(N) = \tau \langle\!\langle {I_{\mathrm{QD}}}^2 \rangle\!\rangle$, with $\langle\!\langle {I_{\mathrm{QD}}}^2 \rangle\!\rangle$ being the diffusion coefficient~\cite{Landi2024}. Given this, the asymptotic Fisher information for such a current measurement can be expressed as $  F(T_{\mathrm{ph}}) = \tau (\partial_{T_{\mathrm{ph}}}\langle {I_\mathrm{QD}}\rangle)^2/\langle\!\langle {I_{\mathrm{QD}}}^2 \rangle\!\rangle$~\cite{Khandelwal2025}.

In our case, the average current reads $\langle I_{\mathrm{QD}}\rangle = \Gamma (f-f_R)/2$. As we assume that the finite reservoir is large enough, i.e., $\zeta \gg 1$, we neglect temperature fluctuations in the stationary state. In this way, the calculation of the diffusion coefficient gives (see the SI~\cite{SI} for details)
\begin{equation}
    \langle\!\langle {I_{\mathrm{QD}}}^2 \rangle\!\rangle = \Gamma\left[  \frac{f\bar{f}+f_R\bar{f}_R}{2} + \frac{(f_R - f)^2}{4}\right].
\end{equation}
Using~\eqref{deltas}, we obtain the following expression for the Fisher information,
\begin{equation}\label{F_IIApprox}
    F_{II}(T_{\mathrm{ph}}) \approx \chi_e \chi_{\mathrm{ph}} \frac{\Gamma \tau}{(f+f_R)(\bar{f} + \bar{f}_R)}.
\end{equation}
We can immediately see that the first two factors in Eq.~\eqref{F_IIApprox}, $\chi_e$ and $\chi_{\mathrm{ph}}$, 
are the same as in Eq.~\eqref{ApproxF_I}.
The third factor in Eq.~\eqref{F_IIApprox} quantifies how much information about the phonon temperature can be extracted from the total current flowing through the QD.

\begin{figure}
    \includegraphics[width=0.99\linewidth]{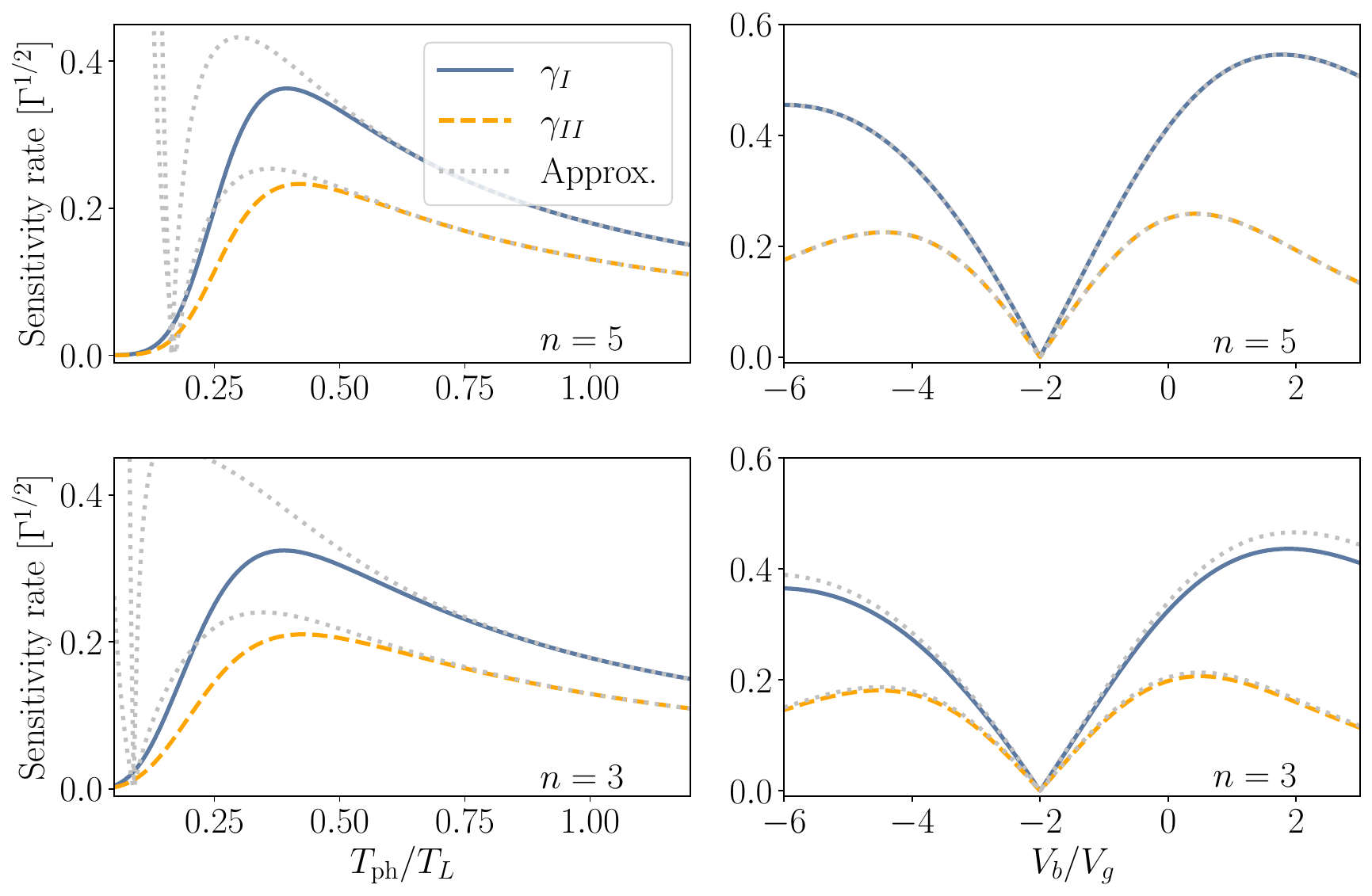}
    \caption{\label{Fig2} Sensitivity rate of the three thermometry schemes as a function of: (left column) phonon temperature and (right column) bias voltage. The top row has electron-phonon coupling with an exponent $n = 5$, while the bottom row has $n=3$. The parameters used are $T_L = T_R = 10\Gamma$, $V_b = 40\Gamma$, $\sigma = 300\Gamma/T_L$, and $V_g = -10\Gamma$. When $n=5$, we use $\Sigma = 1000\sigma/T_L^3$, and when $n=3$, we use $\Sigma = 100\sigma/T_L$. In the right column, we set the phonon temperature to $T_\mathrm{ph} = 8\Gamma$.}
\end{figure}

\textit{Strategy III}.---
Lastly, we consider \emph{repeated measurements} on the system. The protocol is as follows. After the QD evolve into the NESS, a measurement of the QD occupation is performed. 
Then, the system is again allowed to evolve for a sufficiently long time, so that it relaxes back into the NESS, and a new measurement of the QD occupation is realized.
This procedure is repeated $N$ times, thereby giving rise to $N$ independent measurement outcomes.
In this context, the Fisher information for the measurement of the QD occupation can be expressed as $F_{III}(T_{\mathrm{ph}}) = (\partial_{T_\mathrm{ph}}\langle\hat{n}\rangle)^2/\langle\!\langle \hat{n}^2 \rangle\!\rangle$, where $\langle \hat{n}\rangle = p_1^{\mathrm{ss}}$ is the average occupation of the QD, while its fluctuations are given by
\begin{equation}
    \langle\!\langle \hat{n}^2 \rangle\!\rangle = \langle \hat{n}^2 \rangle - \langle \hat{n} \rangle^2 = \frac{f \bar{f} + f_R \bar{f}_R}{2} + \frac{(f_R - f)^2}{4},
\end{equation}
where we used that $\langle \hat{n}^2 \rangle = \langle \hat{n}\rangle$. Using~\eqref{deltas}, we get 
\begin{equation}\label{F_III}
    F_{III}(T_{\mathrm{ph}}) \approx  
    \chi_e \chi_{\mathrm{ph}}
    \frac{1}{(f+f_R)(\bar{f} + \bar{f}_R)}.
\end{equation}

We note that the Fisher information for a single occupation measurement in Eq.~\eqref{F_III} only differs from $F_{II}$ in Eq.~\eqref{F_IIApprox} by a factor of $\Gamma \tau$. If we now express $N =\tau/\tau_M$, where $\tau_M$ is the time interval between measurements, we have that
\begin{equation}
    N F_{III} = \frac{F_{II}}{\Gamma \tau_M}.
\end{equation}
In order to ensure that the system has reached its NESS before performing a new measurement of the occupation, $\tau_M$, must be significantly larger than the relaxation timescale of the system, $\tau_{\mathrm{rel}}$. The latter is determined by the non-zero eigenvalue of the rate matrix $\mathcal{W}$ in Eq.~\eqref{RateEq}, and is given by $\tau_{\mathrm{rel}} = 1/2\Gamma$. Thus, if we fix the time between measurements to be $\tau_M = 2 \eta \tau_{\rm rel}$ for some constant factor $\eta > 1 $, then we find $NF_{III} = F_{II}/\eta$, i.e.~the total Fisher information for strategy $\emph{III}$ is less than for strategy \emph{II}. Finally, we note that the Fisher information in strategies \emph{II} and \emph{III} are only proportional if the coupling strength $\Gamma$ with the left and right reservoirs are the same.

\begin{figure}
\includegraphics[width=0.89\columnwidth]{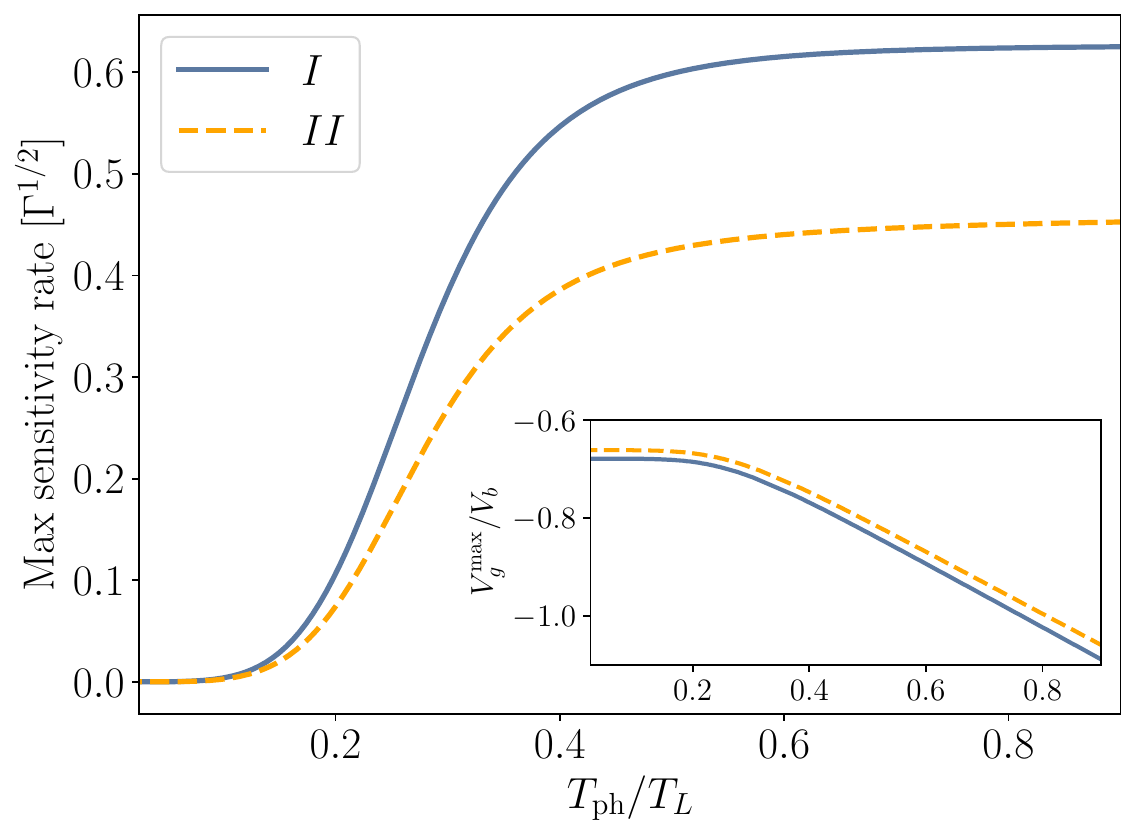}
\caption{\label{Fig3} 
Optimised sensitivity rate over gate voltage as the phonon temperature $T_\mathrm{ph}$ is varied. The inset shows the corresponding optimal bias voltage. The parameters used are $n=5$, $T_L = T_R = 10\Gamma$, $V_b = 40\Gamma$, $\sigma = 300\Gamma/T_L$, and $\Sigma = 1000\sigma/T_L^3$. For these parameters, we have that $\zeta = 300$, and $\xi\gg 1$, provided the phonon temperature is sufficiently large, i.e., $T_\mathrm{ph} \gtrsim T_L/3$.  }
\end{figure}

\textit{Thermometric precision}.--- The Fisher information in Eqs.~\eqref{ApproxF_I} and~\eqref{F_IIApprox}, increase linearly with time. In order to compare the thermometric performance of the current-based strategies \emph{I} and \emph{II}, we define the sensitivity rate as
\begin{equation}\label{precision}
    \gamma_l^2 =  \frac{T_{\mathrm{ph}}^2 F_l}{\tau},
\end{equation}
where $l =\ $\emph{I}, \emph{II} and $F_l$ is the corresponding Fisher information. We evaluate the sensitivity rates numerically, which are plotted in Fig.~\ref{Fig2} for strategies $I$ (solid blue lines) and $II$ (dashed orange lines) as a function of the phonon temperature, $T_{\mathrm{ph}}/T_L$ (left column) and bias voltage $V_b/V_g$ (right column). They are shown for phonon-electron coupling models 
for which $n = 5$ (top row) and $n = 3$ (bottom row). We also plot the approximations in Eqs.~\eqref{ApproxF_I} and~\eqref{F_IIApprox} (gray dotted lines). In all cases, we see that strategy \emph{I}, where the sequence of jumps is monitored, significantly outperforms strategy \emph{II}. This is expected, since Eq.~\eqref{FIcontinuous} corresponds to the optimal measurement performed on the global system during the time $\tau$.

In the left column of Fig.~\ref{Fig2}, we see that the sensitivity rates display a non-monotonic behaviour as a function of $T_{\mathrm{ph}}$. 
The approximations agree with the numerical results for $T_{\mathrm{ph}}$ large enough, so that $\xi \gg 1$. 
We also note that we get a significant difference in the value of sensitivity rates for different values of $n$. More specifically, the sensitivity rate is always higher for $n = 5$ than for $n = 3$. This can be understood as $\xi \propto T_{\mathrm{ph}}^n$, i.e., phonons couple more strongly to the finite reservoir for $n=5$ than for $n=3$. In the right column of Fig.~\ref{Fig2}, both $\gamma_I$ and $\gamma_{II}$ vanish at the point where the chemical potential of the finite reservoir is equal to the QD level energy, $\mu = \epsilon$. This is because $\chi_e = 0$ at this point, as the tunneling rates become independent of finite reservoir temperature. We also see that the sensitivity rates increase linearly from the point where $\mu=\epsilon$, in agreement with Eqs.~\eqref{ApproxF_I} and~\eqref{F_IIApprox}.
 Overall, we see that the numerically calculated sensitivities agree with the approximations to a remarkable degree in the considered range for the voltage bias.

As we have seen, Fig.~\ref{Fig2} demonstrates the existence of optimal current-based sensing of phonon temperature for any set of parameters. We further numerically investigate this by considering the optimisation of sensitivity rates over the experimentally tunable gate voltage, $V_g$,
\begin{equation}
    \gamma_l^{\mathrm{max}} = \max_{V_g} \gamma_l,
\end{equation}
for a fixed value of phonon temperature. We denote the gate voltage that maximizes the sensitivity rate by $V_g^{\mathrm{max}}$. The result of the optimisation is shown in Fig.~\ref{Fig3} for both strategies \emph{I} and \emph{II}. We see that, for $T_{\mathrm{ph}} \gtrsim T_L/3$, a very favorable and robust estimation regime is achieved. The inset shows the corresponding optimal gate voltage for both strategies. These results can guide experimentalists to implement optimal phonon temperature sensing schemes by tuning the gate voltage.

\textit{Conclusion}.--- We introduced an approach harnessing finite reservoirs for current-based phonon temperature estimation in the long time limit. By showing that the Fisher information for jump trajectories corresponds to the quantum Fisher information for any classical rate equation, we were able to demonstrate that monitoring the sequence of electron jumps between the QD system and the finite reservoir (strategy \emph{I}) enables the best possible current-based sensitivity. 
We also investigated the sensitivities of measuring the total charge transferred (strategy \emph{II}) and the QD occupation (strategy \emph{III}). For large but still finite reservoirs, we derived analytical expressions for the Fisher information in strategies \emph{I}, \emph{II}  and \emph{III}. These expressions clearly capture the finite reservoir's thermometric contribution via common multiplicative factors.
Furthermore, we demonstrate optimal precision within current strategies \emph{I} and \emph{II} by tuning the gate voltage.

In most experimental scenarios, it will be easiest to use the estimation scheme \emph{II}, which requires only a simple DC current measurement. Strategy \emph{III} can be realized using fast real-time detection of the QD charge state, e.g., via radio-frequency reflectometry~\cite{Vigneau2023}. However, as we have shown, this does not give any significant advantage in our setting, since strategy \emph{II} yields a higher rate of precision gain once the time between measurements in strategy $\emph{III}$ is accounted for. Strategy \emph{I} is theoretically optimal but also the most challenging to realize, as it requires resolving each individual exchange of charge between the QD and the finite reservoir. Such detection of jumps mediated by a specific reservoir is not out of the question in more complex nonequilibrium setups, e.g.~as demonstrated recently using real-time charge detection on a double-quantum-dot system in the high-bias, Coulomb blockade regime~\cite{Wadhia2025a}.

In conclusion, our results show how finite reservoirs can be harnessed for metrology in nonequilibrium steady states. While we focus on an electronic transport setup, our results serve as a basis for the investigation of how finite reservoirs can be harnessed for current-based parameter estimation in other scenarios, such as superconducting qubits coupled to bosonic environments~\cite{Spiecker2023, Spiecker2024}. 

\textit{Acknowledgments}. 
We gratefully acknowledge discussions with Gabriel T. Landi regarding the QFI for classical rate equations. We also thank Luca Banszerus for insightful discussions. S.B. and M.T.M. acknowledge the financial support of the Royal
Society and Research Ireland. M.T.M. is supported
by a Royal Society University Research Fellowship. S.M. acknowledges financial support from Provincia Autonoma di Trento (PAT); from the Q@TN Initiative; from the National Quantum Science and Technology Institute through the PNRR MUR project under Grant PE0000023-NQSTI, co-funded by the European Union – NextGeneration EU.
This project is co-funded by the
European Union (Quantum Flagship project ASPECTS, Grant
Agreement No. 101080167) and UK Research \& Innovation
(UKRI). Views and opinions expressed are however those of
the authors only and do not necessarily reflect those of the
European Union, Research Executive Agency or UKRI. Neither the European Union nor UKRI can be held responsible
for them.

\bibliography{reference.bib}

@article{Wadhia2025a,
  title = {Entropic {{Costs}} of {{Extracting Classical Ticks}} from a {{Quantum Clock}}},
  author = {Wadhia, Vivek and Meier, Florian and Fedele, Federico and Silva, Ralph and Nurgalieva, Nuriya and Craig, David L. and Jirovec, Daniel and {Saez-Mollejo}, Jaime and Ballabio, Andrea and Chrastina, Daniel and Isella, Giovanni and Huber, Marcus and Mitchison, Mark T. and Erker, Paul and Ares, Natalia},
  year = 2025,
  month = nov,
  journal = {Physical Review Letters},
  volume = {135},
  number = {20},
  pages = {200407},
  issn = {0031-9007, 1079-7114},
  doi = {10.1103/5rtj-djfk},
  urldate = {2026-04-30},
  langid = {english}
}

@article{Vigneau2023,
  title = {Probing Quantum Devices with Radio-Frequency Reflectometry},
  author = {Vigneau, Florian and Fedele, Federico and Chatterjee, Anasua and Reilly, David and Kuemmeth, Ferdinand and {Gonzalez-Zalba}, M. Fernando and Laird, Edward and Ares, Natalia},
  year = 2023,
  month = feb,
  journal = {Appl. Phys. Rev.},
  volume = {10},
  number = {2},
  pages = {021305},
  issn = {1931-9401},
  doi = {10.1063/5.0088229},
  urldate = {2023-07-03}
}

@book{Kay2013,
  title = {Fundamentals of Statistical Signal Processing. 1: {{Estimation}} Theory},
  shorttitle = {Fundamentals of Statistical Signal Processing. 1},
  author = {Kay, Steven M.},
  year = 2013,
  edition = {20. pr},
  publisher = {Prentice Hall PTR},
  address = {Upper Saddle River, NJ},
  isbn = {978-0-13-345711-7},
  langid = {english}
}

@article{PARIS_quantum_estimation,
author = {Paris, MATTEO G. A.},
title = {QUANTUM ESTIMATION FOR QUANTUM TECHNOLOGY},
journal = {Int. J. Quantum Inf.},
volume = {07},
number = {supp01},
pages = {125-137},
year = {2009},
doi = {10.1142/S0219749909004839}
}

@article{singh_detailed_2013,
	title = {Detailed consideration of the electron-phonon thermal conductance at metal-dielectric interfaces},
	volume = {102},
	issn = {0003-6951},
	url = {https://doi.org/10.1063/1.4804383},
	doi = {10.1063/1.4804383},
	number = {18},
	journal = {Appl. Phys. Lett.},
	author = {Singh, Piyush and Seong, Myunghoon and Sinha, Sanjiv},
	month = may,
	year = {2013},
	pages = {181906},
}

@article{wiesner_electronphonon_2022,
	title = {Electron–phonon coupling in copper intercalated {B}i$_2${S}e$_3$},
	volume = {12},
	issn = {2045-2322},
	url = {https://doi.org/10.1038/s41598-022-15909-w},
	doi = {10.1038/s41598-022-15909-w},
	journal = {Sci. Rep.},
	author = {Wiesner, Maciej and Koski, Kristie and Laitinen, Antti and Manninen, Juuso and Zyuzin, Alexander A. and Hakonen, Pertti},
	month = jul,
	year = {2022},
	pages = {12097},
}

@misc{mihailescu2025,
      title={Quantum Sensing with Nanoelectronics: Fisher Information for an Applied Perturbation}, 
      author={George Mihailescu and Anthony Kiely and Andrew K. Mitchell},
      year={2025},
      eprint={2406.18662},
      archivePrefix={arXiv},
      primaryClass={quant-ph},
      url={https://arxiv.org/abs/2406.18662}, 
}

@article{Khandelwal2025,
  title = {Current-based metrology with two-terminal mesoscopic conductors},
  author = {Khandelwal, Shishir and Landi, Gabriel T. and Haack, G\'eraldine and Mitchison, Mark T.},
  journal = {Phys. Rev. B},
  volume = {112},
  issue = {16},
  pages = {L161409},
  numpages = {7},
  year = {2025},
  month = {Oct},
  publisher = {American Physical Society},
  doi = {10.1103/cpp5-5yw4},
  url = {https://link.aps.org/doi/10.1103/cpp5-5yw4}
}

@article{Matern2024,
  title = {Thermoelectric cooling of a finite reservoir coupled to a quantum dot},
  author = {Matern, Stephanie and Moreira, Saulo V. and Samuelsson, Peter and Leijnse, Martin},
  journal = {Phys. Rev. B},
  volume = {110},
  issue = {20},
  pages = {205423},
  numpages = {9},
  year = {2024},
  month = {Nov},
  publisher = {American Physical Society},
  doi = {10.1103/PhysRevB.110.205423},
  url = {https://link.aps.org/doi/10.1103/PhysRevB.110.205423}
}

@article{pekola_quantum_2018,
	title = {Quantum {Noise} of {Electron}–{Phonon} {Heat} {Current}},
	volume = {191},
	issn = {1573-7357},
	url = {https://doi.org/10.1007/s10909-018-1854-y},
	doi = {10.1007/s10909-018-1854-y},
	number = {5},
	journal = {J. Low Temp. Phys.},
	author = {Pekola, Jukka P. and Karimi, Bayan},
	month = jun,
	year = {2018},
	pages = {373--379},
}

@article{Chow1996,
  title = {Phonon Emission from a 2{D} Electron Gas: Evidence of Transition to the Hydrodynamic Regime},
  author = {Chow, Edmond and Wei, H. P. and Girvin, S. M. and Shayegan, M.},
  journal = {Phys. Rev. Lett.},
  volume = {77},
  issue = {6},
  pages = {1143--1146},
  numpages = {0},
  year = {1996},
  month = {Aug},
  publisher = {American Physical Society},
  doi = {10.1103/PhysRevLett.77.1143},
  url = {https://link.aps.org/doi/10.1103/PhysRevLett.77.1143}
}

@article{Schmidt2004,
  title = {Photon-Mediated Thermal Relaxation of Electrons in Nanostructures},
  author = {Schmidt, D. R. and Schoelkopf, R. J. and Cleland, A. N.},
  journal = {Phys. Rev. Lett.},
  volume = {93},
  issue = {4},
  pages = {045901},
  numpages = {4},
  year = {2004},
  month = {Jul},
  publisher = {American Physical Society},
  doi = {10.1103/PhysRevLett.93.045901},
  url = {https://link.aps.org/doi/10.1103/PhysRevLett.93.045901}
}

@article{Giazotto2006,
  title = {Opportunities for mesoscopics in thermometry and refrigeration: Physics and applications},
  author = {Giazotto, Francesco and Heikkil\"a, Tero T. and Luukanen, Arttu and Savin, Alexander M. and Pekola, Jukka P.},
  journal = {Rev. Mod. Phys.},
  volume = {78},
  issue = {1},
  pages = {217--274},
  numpages = {0},
  year = {2006},
  month = {Mar},
  publisher = {American Physical Society},
  doi = {10.1103/RevModPhys.78.217},
  url = {https://link.aps.org/doi/10.1103/RevModPhys.78.217}
}

@article{Pothier1997,
  title = {Energy Distribution Function of Quasiparticles in Mesoscopic Wires},
  author = {Pothier, H. and Gu\'eron, S. and Birge, Norman O. and Esteve, D. and Devoret, M. H.},
  journal = {Phys. Rev. Lett.},
  volume = {79},
  issue = {18},
  pages = {3490--3493},
  numpages = {0},
  year = {1997},
  month = {Nov},
  publisher = {American Physical Society},
  doi = {10.1103/PhysRevLett.79.3490},
  url = {https://link.aps.org/doi/10.1103/PhysRevLett.79.3490}
}

@article{Maradan2014,
author = {Maradan, D and Casparis, L and Liu, T.-M. and Biesinger, D E F and Scheller, C P and Zumb{\"{u}}hl, D M and Zimmerman, J D and Gossard, A C},
doi = {10.1007/s10909-014-1169-6},
issn = {1573-7357},
journal = {J. Low Temp. Phys.},
number = {5},
pages = {784--798},
title = {{GaAs Quantum Dot Thermometry Using Direct Transport and Charge Sensing}},
url = {https://doi.org/10.1007/s10909-014-1169-6},
volume = {175},
year = {2014}
}

@article{Hofer2017,
  title = {Quantum Thermal Machine as a Thermometer},
  author = {Hofer, Patrick P. and Brask, Jonatan Bohr and Perarnau-Llobet, Mart\'{\i} and Brunner, Nicolas},
  journal = {Phys. Rev. Lett.},
  volume = {119},
  issue = {9},
  pages = {090603},
  numpages = {7},
  year = {2017},
  month = {Sep},
  publisher = {American Physical Society},
  doi = {10.1103/PhysRevLett.119.090603},
  url = {https://link.aps.org/doi/10.1103/PhysRevLett.119.090603}
}

@article{Yang2019,
  title = {Thermal transistor and thermometer based on Coulomb-coupled conductors},
  author = {Yang, Jing and Elouard, Cyril and Splettstoesser, Janine and Sothmann, Bj\"orn and S\'anchez, Rafael and Jordan, Andrew N.},
  journal = {Phys. Rev. B},
  volume = {100},
  issue = {4},
  pages = {045418},
  numpages = {10},
  year = {2019},
  month = {Jul},
  publisher = {American Physical Society},
  doi = {10.1103/PhysRevB.100.045418},
  url = {https://link.aps.org/doi/10.1103/PhysRevB.100.045418}
}

@article{Boeyens2023,
	author = {Boeyens, Julia and Annby-Andersson, Bj{\"o}rn and Bakhshinezhad, Pharnam and Haack, G{\'e}raldine and Perarnau-Llobet, Mart{\'\i} and Nimmrichter, Stefan and Potts, Patrick P and Mehboudi, Mohammad},
	doi = {10.1088/1367-2630/ad0e8a},
	journal = {New J. Phys.},
	month = {dec},
	number = {12},
	pages = {123009},
	publisher = {IOP Publishing},
	title = {Probe thermometry with continuous measurements},
	url = {https://doi.org/10.1088/1367-2630/ad0e8a},
	volume = {25},
	year = {2023}}

@article{Brange2018,
  title = {Nanoscale quantum calorimetry with electronic temperature fluctuations},
  author = {Brange, F. and Samuelsson, P. and Karimi, B. and Pekola, J. P.},
  journal = {Phys. Rev. B},
  volume = {98},
  issue = {20},
  pages = {205414},
  numpages = {10},
  year = {2018},
  month = {Nov},
  publisher = {American Physical Society},
  doi = {10.1103/PhysRevB.98.205414},
  url = {https://link.aps.org/doi/10.1103/PhysRevB.98.205414}
}

@article{Karimi2020,
	author = {Karimi, Bayan and Brange, Fredrik and Samuelsson, Peter and Pekola, Jukka P.},
	date = {2020/01/17},
	doi = {10.1038/s41467-019-14247-2},
	id = {Karimi2020},
	isbn = {2041-1723},
	journal = {Nat. Comm.},
	number = {1},
	pages = {367},
	title = {Reaching the ultimate energy resolution of a quantum detector},
	url = {https://doi.org/10.1038/s41467-019-14247-2},
	volume = {11},
	year = {2020}}

@article{Moreira2023,
  title = {Stochastic Thermodynamics of a Quantum Dot Coupled to a Finite-Size Reservoir},
  author = {Moreira, Saulo V. and Samuelsson, Peter and Potts, Patrick P.},
  journal = {Phys. Rev. Lett.},
  volume = {131},
  issue = {22},
  pages = {220405},
  numpages = {7},
  year = {2023},
  month = {Dec},
  publisher = {American Physical Society},
  doi = {10.1103/PhysRevLett.131.220405},
  url = {https://link.aps.org/doi/10.1103/PhysRevLett.131.220405}
}

@article{Mamede2026,
  title = {Steady-state heat engines driven by finite reservoirs},
  author = {Mamede, Iago N. and Moreira, Saulo V. and Mitchison, Mark T. and Fiore, Carlos E.},
  journal = {Phys. Rev. E},
  volume = {113},
  issue = {1},
  pages = {014115},
  numpages = {13},
  year = {2026},
  month = {Jan},
  publisher = {American Physical Society},
  doi = {10.1103/hwj9-q7hf},
  url = {https://link.aps.org/doi/10.1103/hwj9-q7hf}
}

@article{Yuan2026,
  title = {Optimizing thermodynamic cycles with two finite-sized reservoirs},
  author = {Yuan, Hong and Ma, Yu-Han and Sun, C. P.},
  journal = {Phys. Rev. E},
  volume = {105},
  issue = {2},
  pages = {L022101},
  numpages = {5},
  year = {2022},
  month = {Feb},
  publisher = {American Physical Society},
  doi = {10.1103/PhysRevE.105.L022101},
  url = {https://link.aps.org/doi/10.1103/PhysRevE.105.L022101}
}

@incollection{Ingold1992,
  author    = {Ingold, G.-L. and Nazarov, Y. V.},
  title     = {Charge Tunneling Rates in Ultrasmall Junctions},
  booktitle = {Single Charge Tunneling},
  editor    = {Grabert, H. and Devoret, M. H.},
  chapter   = {2},
  pages     = {21--107},
  publisher = {Springer},
  address   = {New York},
  year      = {1992}
}

@article{Gammelmark2014,
  title = {Fisher Information and the Quantum Cram\'er-Rao Sensitivity Limit of Continuous Measurements},
  author = {Gammelmark, S\o{}ren and M\o{}lmer, Klaus},
  journal = {Phys. Rev. Lett.},
  volume = {112},
  issue = {17},
  pages = {170401},
  numpages = {5},
  year = {2014},
  month = {Apr},
  publisher = {American Physical Society},
  doi = {10.1103/PhysRevLett.112.170401},
  url = {https://link.aps.org/doi/10.1103/PhysRevLett.112.170401}
}

@article{Esposito2009,
  title = {Nonequilibrium fluctuations, fluctuation theorems, and counting statistics in quantum systems},
  author = {Esposito, Massimiliano and Harbola, Upendra and Mukamel, Shaul},
  journal = {Rev. Mod. Phys.},
  volume = {81},
  issue = {4},
  pages = {1665--1702},
  numpages = {0},
  year = {2009},
  month = {Dec},
  publisher = {American Physical Society},
  doi = {10.1103/RevModPhys.81.1665},
  url = {https://link.aps.org/doi/10.1103/RevModPhys.81.1665}
}

@article{Landi2024,
  title = {Current Fluctuations in Open Quantum Systems: Bridging the Gap Between Quantum Continuous Measurements and Full Counting Statistics},
  author = {Landi, Gabriel T. and Kewming, Michael J. and Mitchison, Mark T. and Potts, Patrick P.},
  journal = {PRX Quantum},
  volume = {5},
  issue = {2},
  pages = {020201},
  numpages = {86},
  year = {2024},
  month = {Apr},
  publisher = {American Physical Society},
  doi = {10.1103/PRXQuantum.5.020201},
  url = {https://link.aps.org/doi/10.1103/PRXQuantum.5.020201}
}

@article{Smiga2023,
  title = {Stochastic metrology and the empirical distribution},
  author = {Smiga, Joseph A. and Radaelli, Marco and Binder, Felix C. and Landi, Gabriel T.},
  journal = {Phys. Rev. Res.},
  volume = {5},
  issue = {3},
  pages = {033150},
  numpages = {14},
  year = {2023},
  month = {Sep},
  publisher = {American Physical Society},
  doi = {10.1103/PhysRevResearch.5.033150},
  url = {https://link.aps.org/doi/10.1103/PhysRevResearch.5.033150}
}

@article{Beechem2007,
	author = {Beechem, Thomas and Graham, Samuel and Kearney, Sean P. and Phinney, Leslie M. and Serrano, Justin R.},
	doi = {10.1063/1.2738946},
	issn = {0034-6748},
	journal = {Rev. Sci. Instrum.},
	month = {06},
	number = {6},
	pages = {061301},
	title = {Invited Article: Simultaneous mapping of temperature and stress in microdevices using micro-Raman spectroscopy},
	url = {https://doi.org/10.1063/1.2738946},
	volume = {78},
	year = {2007}}

@article{Beechem2015,
	author = {Beechem, Thomas and Yates, Luke and Graham, Samuel},
	doi = {10.1063/1.4918623},
	issn = {0034-6748},
	journal = {Rev. Sci. Instrum.},
	month = {04},
	number = {4},
	pages = {041101},
	title = {Invited Review Article: Error and uncertainty in Raman thermal conductivity measurements},
	url = {https://doi.org/10.1063/1.4918623},
	volume = {86},
	year = {2015}}

@article{Sandell2020,
	author = {Sandell, Susanne and Ch{\'a}vez-{\'A}ngel, Emigdio and El Sachat, Alexandros and He, Jianying and Sotomayor Torres, Clivia M. and Maire, Jeremie},
	doi = {10.1063/5.0020239},
	issn = {0021-8979},
	journal = {J. Appl. Phys.},
	month = {10},
	number = {13},
	pages = {131101},
	title = {Thermoreflectance techniques and Raman thermometry for thermal property characterization of nanostructures},
	url = {https://doi.org/10.1063/5.0020239},
	volume = {128},
	year = {2020}}

@article{Radaelli2023,
	author = {Radaelli, Marco and Landi, Gabriel T and Modi, Kavan and Binder, Felix C},
	doi = {10.1088/1367-2630/acd321},
	journal = {New J. Phys.},
	month = {jun},
	number = {5},
	pages = {053037},
	publisher = {IOP Publishing},
	title = {Fisher information of correlated stochastic processes},
	url = {https://doi.org/10.1088/1367-2630/acd321},
	volume = {25},
	year = {2023}}

@article{Pekola2021,
  title = {Colloquium: Quantum heat transport in condensed matter systems},
  author = {Pekola, Jukka P. and Karimi, Bayan},
  journal = {Rev. Mod. Phys.},
  volume = {93},
  issue = {4},
  pages = {041001},
  numpages = {25},
  year = {2021},
  month = {Oct},
  publisher = {American Physical Society},
  doi = {10.1103/RevModPhys.93.041001},
  url = {https://link.aps.org/doi/10.1103/RevModPhys.93.041001}
}

@article{Dutta2020,
  title = {Single-Quantum-Dot Heat Valve},
  author = {Dutta, B. and Majidi, D. and Talarico, N. W. and Lo Gullo, N. and Courtois, H. and Winkelmann, C. B.},
  journal = {Phys. Rev. Lett.},
  volume = {125},
  issue = {23},
  pages = {237701},
  numpages = {6},
  year = {2020},
  month = {Dec},
  publisher = {American Physical Society},
  doi = {10.1103/PhysRevLett.125.237701},
  url = {https://link.aps.org/doi/10.1103/PhysRevLett.125.237701}
}

@article{Spiecker2023,
	author = {Spiecker, Martin and Paluch, Patrick and Gosling, Nicolas and Drucker, Niv and Matityahu, Shlomi and Gusenkova, Daria and G{\"u}nzler, Simon and Rieger, Dennis and Takmakov, Ivan and Valenti, Francesco and Winkel, Patrick and Gebauer, Richard and Sander, Oliver and Catelani, Gianluigi and Shnirman, Alexander and Ustinov, Alexey V. and Wernsdorfer, Wolfgang and Cohen, Yonatan and Pop, Ioan M.},
	date = {2023/09/01},
	doi = {10.1038/s41567-023-02082-8},
	id = {Spiecker2023},
	isbn = {1745-2481},
	journal = {Nat. Phys.},
	number = {9},
	pages = {1320--1325},
	title = {Two-level system hyperpolarization using a quantum Szilard engine},
	url = {https://doi.org/10.1038/s41567-023-02082-8},
	volume = {19},
	year = {2023}}

@article{Spiecker2024,
  title = {Solomon equations for qubit and two-level systems: Insights into non-Poissonian quantum jumps},
  author = {Spiecker, Martin and Pavlov, Andrei I. and Shnirman, Alexander and Pop, Ioan M.},
  journal = {Phys. Rev. A},
  volume = {109},
  issue = {5},
  pages = {052218},
  numpages = {24},
  year = {2024},
  month = {May},
  publisher = {American Physical Society},
  doi = {10.1103/PhysRevA.109.052218},
  url = {https://link.aps.org/doi/10.1103/PhysRevA.109.052218}
}

@article{Champain2024,
  title = {Real-time millikelvin thermometry in a semiconductor-qubit architecture},
  author = {Champain, V. and Schmitt, V. and Bertrand, B. and Niebojewski, H. and Maurand, R. and Jehl, X. and Winkelmann, C.B. and De Franceschi, S. and Brun, B.},
  journal = {Phys. Rev. Appl.},
  volume = {21},
  issue = {6},
  pages = {064039},
  numpages = {8},
  year = {2024},
  month = {Jun},
  publisher = {American Physical Society},
  doi = {10.1103/PhysRevApplied.21.064039},
  url = {https://link.aps.org/doi/10.1103/PhysRevApplied.21.064039}
}

@article{Strasberg2021-2,
	author = {Strasberg, P. and D{\'\i}az, M. G. and Riera-Campeny, A.},
    title = {{C}lausius inequality for finite baths reveals universal efficiency improvements},
	doi = {10.1103/PhysRevE.104.L022103},
	issue = {2},
	journal = {Phys. Rev. E},
	pages = {L022103},
	volume = {104},
	year = {2021},
	}

@article{Strasberg2021-1,
	author = {Strasberg, P. and Winter, A.},
    title = {First and Second Law of Quantum Thermodynamics: A Consistent Derivation Based on a Microscopic Definition of Entropy},
	doi = {10.1103/PRXQuantum.2.030202},
	journal = {PRX Quantum},
	pages = {030202},
	volume = {2},
	year = {2021}}

@article{RieraCampeny2022,
	author = {Riera-Campeny, A. and Sanpera, A. and Strasberg, P.},
    title = {Open quantum systems coupled to finite baths: A hierarchy of master equations},
	doi = {10.1103/PhysRevE.105.054119},
	journal = {Phys. Rev. E},
	pages = {054119},
	volume = {105},
	year = {2022},
	}

@article{RieraCampeny2021,
	author = {Riera-Campeny, A. and Sanpera, A. and Strasberg, P.},
    title = {Quantum Systems Correlated with a Finite Bath: Nonequilibrium Dynamics and Thermodynamics},
	doi = {10.1103/PRXQuantum.2.010340},
	journal = {PRX Quantum},
	pages = {010340},
	volume = {2},
	year = {2021}}

@article{Elouard2022,
  title = {Extending the Laws of Thermodynamics for Arbitrary Autonomous Quantum Systems},
  author = {Elouard, C. and Lombard Latune, C.},
  journal = {PRX Quantum},
  volume = {4},
  pages = {020309},
  year = {2023},
  doi = {10.1103/PRXQuantum.4.020309},
   }

@article{Golovach2011,
  title = {Single-dopant resonance in a single-electron transistor},
  author = {Golovach, V. N. and Jehl, X. and Houzet, M. and Pierre, M. and Roche, B. and Sanquer, M. and Glazman, L. I.},
  journal = {Phys. Rev. B},
  volume = {83},
  issue = {7},
  pages = {075401},
  numpages = {13},
  year = {2011},
  month = {Feb},
  publisher = {American Physical Society},
  doi = {10.1103/PhysRevB.83.075401},
  url = {https://link.aps.org/doi/10.1103/PhysRevB.83.075401}
}

@article{Schaller2014,
	author = {Schaller, G. and Nietner, C. and Brandes, T.},
    title = {Relaxation dynamics of meso-reservoirs},
	doi = {10.1088/1367-2630/16/12/125011},
	journal = {New J. Phys.},
	pages = {125011},
	volume = {16},
	year = {2014},
    }

@article{Donvil2022,
	author = {B. Donvil and J. Ankerhold},
	doi = {10.1088/1751-8121/ac677d},
	journal = {J. Phys. A: Math. Theor.},
	pages = {225303},
	publisher = {IOP Publishing},
	title = {Apparent heating due to imperfect calorimetric measurements},
	volume = {55},
	year = {2022},
    }

@article{Grenier2016,
	author = {C. Grenier and C. Kollath and A. Georges},
	doi = {https://doi.org/10.1016/j.crhy.2016.08.013},
    title = {Thermoelectric transport and Peltier cooling of cold atomic gases},
	journal = {C. R. Phys.},
	number = {10},
	pages = {1161},
	volume = {17},
	year = {2016},
	}

@article{Gallego2014,
  title = {Nonequilibrium relaxation transport of ultracold atoms},
  author = {Gallego-Marcos, Fernando and Platero, Gloria and Nietner, Christian and Schaller, Gernot and Brandes, Tobias},
  journal = {Phys. Rev. A},
  volume = {90},
  issue = {3},
  pages = {033614},
  numpages = {12},
  year = {2014},
  month = {Sep},
  publisher = {American Physical Society},
  doi = {10.1103/PhysRevA.90.033614},
  url = {https://link.aps.org/doi/10.1103/PhysRevA.90.033614}
}

@article{Amato2020,
  title = {Noninteracting many-particle quantum transport between finite reservoirs},
  author = {Amato, Giulio and Breuer, Heinz-Peter and Wimberger, Sandro and Rodr\'{\i}guez, Alberto and Buchleitner, Andreas},
  journal = {Phys. Rev. A},
  volume = {102},
  issue = {2},
  pages = {022207},
  numpages = {15},
  year = {2020},
  month = {Aug},
  publisher = {American Physical Society},
  doi = {10.1103/PhysRevA.102.022207},
  url = {https://link.aps.org/doi/10.1103/PhysRevA.102.022207}
}

@book{Cramer1946,
  author    = {Cram{\'e}r, Harald},
  title     = {Mathematical Methods of Statistics},
  year      = {1946},
  publisher = {Princeton University Press},
  address   = {Princeton, NJ}
}

@book{Rao1973,
  author    = {Rao, Calyampudi Radhakrishna},
  title     = {Linear Statistical Inference and Its Applications},
  year      = {1973},
  publisher = {Wiley},
  address   = {New York}
}

@article{blanter_shot_2000,
	title = {Shot noise in mesoscopic conductors},
	volume = {336},
	issn = {0370-1573},
	url = {https://www.sciencedirect.com/science/article/pii/S0370157399001234},
	doi = {10.1016/S0370-1573(99)00123-4},
	number = {1},
	journal = {Phys. Rep.},
	author = {Blanter, Ya. M. and B{\"u}ttiker, M.},
	month = sep,
	year = {2000},
	pages = {1--166},
}

@article{Yamamoto2015,
  title = {Thermodynamics of the mesoscopic thermoelectric heat engine beyond the linear-response regime},
  author = {Yamamoto, Kaoru and Hatano, Naomichi},
  journal = {Phys. Rev. E},
  volume = {92},
  issue = {4},
  pages = {042165},
  numpages = {7},
  year = {2015},
  month = {Oct},
  doi = {10.1103/PhysRevE.92.042165},
  url = {https://link.aps.org/doi/10.1103/PhysRevE.92.042165}
}

@article{Gorini1976,
  author = {Gorini, Vittorio and Kossakowski, Andrzej and Sudarshan, E. C. G.},
  title = {Completely positive dynamical semigroups of {N}-level systems},
  journal = {J. Math. Phys.},
  volume = {17},
  number = {5},
  pages = {821--825},
  year = {1976},
  doi = {10.1063/1.522979}
}

@article{Lindblad1976,
  author = {Lindblad, G\"{o}ran},
  title = {On the generators of quantum dynamical semigroups},
  journal = {Commun. Math. Phys.},
  volume = {48},
  number = {2},
  pages = {119--130},
  year = {1976},
  doi = {10.1007/BF01608499}
}

@misc{SI,
  title ={The {S}upplemental {I}nformation provides derivations of the linear response equations for the transport between the infinite and finite reservoirs, linear response equations for the transport between the infinite and finite reservoirs, and transport expressions between the finite reservoir and the {QD}. {I}t also contains the investigation of the regimes for which $\xi\ll1$ and $\xi\gg1$, the derivation of the quantum {F}isher information for a rate equation in the long time limit, as well as the analytical expressions for current-based {F}isher information in the main text, ${F}_{I}$ and ${F}_{II}$.}
}

@article{Berg2015,
	author = {Berg, Tineke L van den and Brange, Fredrik and Samuelsson, Peter},
	doi = {10.1088/1367-2630/17/7/075012},
	journal = {New J. Phys.},
	month = {jul},
	number = {7},
	pages = {075012},
	publisher = {IOP Publishing},
	title = {Energy and temperature fluctuations in the single electron box},
	url = {https://doi.org/10.1088/1367-2630/17/7/075012},
	volume = {17},
	year = {2015}}

\newpage
\widetext
\begin{center}
	\textbf{\large Supplemental information} 
\end{center}
\setcounter{equation}{0}
\setcounter{figure}{0}
\setcounter{table}{0}
\setcounter{page}{1}
\makeatletter
\renewcommand{\theequation}{S\arabic{equation}}
\renewcommand{\thefigure}{S\arabic{figure}}

\setcounter{equation}{0}
\setcounter{figure}{0}
\setcounter{table}{0}
\setcounter{page}{1}
\setcounter{section}{0}
\makeatletter
\renewcommand{\theequation}{S\arabic{equation}}
\renewcommand{\thefigure}{S\arabic{figure}}

In this supplemental information, we provide derivations and motivations for different concepts introduced in the main text. In Section~\ref{app:linres_inf}, we derive linear response equations for the transport between the infinite and finite reservoirs. Section~\ref{app:linres_QD} derives the transport expressions between the finite reservoir and QD. In Sections~\ref{app:linres_big_sigma} and~\ref{app:linres_big_Sigma}, we investigate the regimes for which $\xi\ll1$ and $\xi\gg1$, respectively. Section~\ref{app:qfi} derives the quantum Fisher information for a rate equation in the long time limit, while in Sections~\ref{app:F_I} and~\ref{appx:F_II} we derive the analytical expressions for current-based Fisher information in the main text, $F_I$ and $F_{II}$.

\section{Linear response transport between infinite and finite reservoir}
\label{app:linres_inf}
We start by deriving linear response equations for the transport between the infinite and finite reservoirs in Eq.~\eqref{eq:inf_currents}. We assume that the transport is ballistic, energy independent, and that the finite reservoir is in quasi-equilibrium, so that we can use the Landauer-B{\"u}tikker formalism to write the average particle current $\expval{I_\mathrm{inf}}$, energy current $\expval{J_\mathrm{inf}}$, and heat current $\expval*{\dot{Q}_\mathrm{
inf}}$, as~\cite{blanter_shot_2000, Yamamoto2015}: 
\begin{align}
    \expval{I_\mathrm{inf}} &=\frac{1}{h} \int \dd E \mathcal{T}(E)\left[f(E, \mu_L,T_L) - f(E,\mu,T\right]\\
    \expval{J_\mathrm{inf}}&=\frac{1}{h}\int \dd E \mathcal{T}(E)(E)\left[f(E, \mu_L,T_L) - f(E,\mu,T\right]\\
    \expval*{\dot{Q}_\mathrm{inf}}&= \expval{J_\mathrm{inf}} - \mu \expval{I_\mathrm{inf}},
\end{align}
where $h$ is the Planck constant. The function $\mathcal{T}(E)$ is called the transmission function, and for the scenario we describe here, it takes the form
\begin{align}
    \mathcal{T}(E) = N\left[\Theta(E - \mu_L -\delta ) - \Theta(E - \mu_L +\delta )\right],
\end{align}
i.e., the transmission is centred around the chemical potential of the infinite reservoir, and has a width of $2\delta$, and $N$ open channels through which the electrons can go. These integrals can be exactly evaluated, yielding 
\begin{align}
    \expval{I_\mathrm{inf}} &= \frac{N}{h}\left(T\left[\ln(1+e^{-\frac{2\delta - \Delta\mu}{2T}}) -\ln(1+e^{\frac{2\delta + \Delta\mu}{2T}}) \right] +T_L \left[\ln(1+e^{\frac{2\delta - \Delta\mu}{2T_L}}) - \ln(1+e^{-\frac{2\delta+\Delta\mu}{2T_L}}) \right] \right), \label{app:I_inf}\\
    \expval{J_\mathrm{inf}} &= \frac{N}{h}\bigg([\delta-\mu_L]\left[T\ln(1+e^{\frac{2\delta-\Delta\mu}{2T}}) -T_L\ln(1+e^{\frac{2\delta+\Delta\mu}{2T}})\right] \label{app:Jinf} \\
    &+ [\delta + \mu_L]\left[T\ln(1+e^{-\frac{2\delta+\Delta\mu}{2T}}) -T_L\ln(1+e^{-\frac{2\delta-\Delta\mu}{2T}}) \right] \nonumber\\
    &+T_L^2\left[\mathrm{Li}_2\left(-e^{-\frac{2\delta-\Delta\mu}{2T_L}}\right) - \mathrm{Li}_2\left(-e^{\frac{2\delta+\Delta\mu}{2T_L}}\right)\right] + T^2\left[\mathrm{Li}_2\left(-e^{\frac{2\delta-\Delta\mu}{2T}}\right) - \mathrm{Li}_2\left(-e^{-\frac{2\delta+\Delta\mu}{2T}}\right)\right]  \bigg)\nonumber,
\end{align}
where $\Delta\mu = \mu_L-\mu$ and $\mathrm{Li}_2(x)$ denotes the dilogarithm. These expressions allow us to numerically solve the equations of motion in the main text in the steady state.

We are now interested in the transport in the linear response regime, i.e. when $N$ and $\delta$ are large, and the change in chemical potential $\Delta\mu$ and temperature $\Delta T = T_L-T$ are small. Taylor expanding Eq. (\ref{app:I_inf}) to first order in $\Delta\mu$ and $\Delta T$ yields 
\begin{align}
    \expval{I_\mathrm{inf}} \approx \frac{N\tanh\left(\frac{\delta}{2\Bar{T}}\right)\Delta\mu}{h} \equiv \sigma\Delta\mu,
\end{align}
where we have identified the linear response electrical conductance $\sigma =N\tanh\left(\frac{\delta}{2\Bar{T}}\right)/h$. Note that due to symmetry the Seebeck coefficient is zero. Similarly, we expand Eq.~(\ref{app:Jinf}), and in the limit of a wide transmission function, we recover
\begin{align}
    \expval*{\dot{Q}_\mathrm{inf}} \approx \frac{\pi^2\Bar{T}\sigma}{3}\Delta T +\Bar{\mu} \sigma \Delta \mu,
\end{align}
and we can immediately see that we recover the electronic thermal conductance $\kappa_\mathrm{el} =\pi^2\Bar{T}\sigma/3$ and Eq.~\eqref{eq:inf_currents} from the main text. 

\section{Linear response transport through QD}
\label{app:linres_QD}
Next, we consider the strict linear response regime where there is a small bias voltage $V_b$ applied between the two infinite reservoirs such that $\mu_L = V_b/2 = - \mu_R$. Furthermore, the temperature of the two infinite reservoirs is taken to be the same, $T_R=T_L$. In this case, the chemical potential and temperature of the finite reservoir is given by $\mu = V_b/2 + \Delta \mu$ and $T = T_L + \Delta T$. 

In the NESS, the average current from the finite reservoir to the QD is the same as the average current from the QD into the right infinite reservoir, and is given by
\begin{align}
\label{app:Ir}
    \expval{I_R} = \Gamma [p_1^{ss} (1-f_R) + p_0^{ss}f_R],
\end{align}
and the steady state occupation of the dot is given by $p_1^{ss} = (f +f_R)/2$, the average of the Fermi functions of the finite reservoir and right infinite reservoir evaluated at the energy of the QD. In the strict linear response regime, the current is given by 
\begin{align}
    \expval{I_R} = G\delta\mu - SG\delta T,
\end{align}
where $\delta \mu = \mu - \mu_R = V_b + \Delta\mu$ and $\delta T = T - T_R = \Delta T$. By expanding Eq.~(\ref{app:Ir}) to first order in $V_b$, $\Delta\mu$ and $\Delta T$, we can read off the conductivity
\begin{align}
    G = \frac{\Gamma}{8T_L}\sech^2\left(\frac{\epsilon}{2T_L}\right),
\end{align}
and the Seebeck coefficient 
\begin{align}
    S  = -\frac{\epsilon}{T_L}
\end{align}
for the transport through the QD. 

\section{Linear response when strongly coupled to infinite reservoir}
\label{app:linres_big_sigma}
Here, we investigate the regime where $\zeta \gg 1$, i.e., the finite reservoir is strongly coupled to the infinite electronic reservoir and weakly coupled to the QD, and $\xi\ll1$, which corresponds to weak coupling between the finite reservoir and phonon bath. 
In this regime, we expect the stationary temperature and chemical potential of the finite reservoir to be very close to those of the infinite reservoir, i.e., $\mu = \mu_L + \Delta\mu$ and $T = T_L + \Delta T_L$. Setting Eqs.~\eqref{MuDot} and~\eqref{TDot} from the main text to zero, and linearising in $\Delta T_L$ and $\Delta \mu$, we can solve for the shifts in temperature and chemical potential, and we find that:
\begin{align}
    \Delta\mu &= -G\frac{V_b(\pi^2T_L\frac{\sigma}{\Sigma} + 3nT_L^{n-1}) + 3S(T_L^n - T_\mathrm{ph}^n)}{(G+\sigma)(\pi^2T_L\frac{\sigma}{\Sigma} +3nT_L^{n-1}) +3GS^2T_L\frac{\sigma}{\Sigma}} \\
    \Delta T_L &= \frac{3(G + \sigma)(T_\mathrm{ph}^n-T_L^n)-3GV_b\epsilon \frac{\sigma}{\Sigma}}{(G+\sigma)(\pi^2T_L^2\frac{\sigma}{\Sigma} +3nT_L)+ 3GS^2T_L\frac{\sigma}{\Sigma}}.
\end{align}
These expressions show a complicated competition between all the currents going in and out of the finite reservoir. To gain further insight, we expand these expressions to first order in $\xi\ll1$,
\begin{align}
    \Delta\mu &\approx -\frac{V_b}{\sigma}G \\
    \Delta T &\approx \frac{3\xi T_L}{\pi^2}\left(1 - \frac{T_L^n}{T_\mathrm{ph}^n} \right) + \frac{3V_b}{\pi^2\sigma}SG.\label{app:delta_T_L}
\end{align}
We clearly see that, in this regime, the chemical potential is set by the competition between the coupling strengths of the finite reservoir to the infinite reservoir and QD, and that the chemical potential is insensitive to the phonon bath coupling. The change in temperature, on the other hand, has two terms. The first term describes the competition between coupling to the infinite reservoir and the phonon bath, while the second term describes the competition between the infinite reservoir and QD coupling. 

Furthermore, we note that in the limit of $\xi \rightarrow 0$ and $\zeta\rightarrow \infty$, i.e., when the coupling to the infinite reservoir is very strong, both $\Delta T$ and $\Delta \mu$ go to zero. As discussed in Ref.~\cite{Matern2024}, this corresponds to the limit where the finiteness of the reservoir becomes irrelevant, and the finite reservoir becomes completely insensitive to the coupling to the phonon bath.

\section{Linear response when strongly coupled to phonon bath}
\label{app:linres_big_Sigma}
\begin{figure}
    \centering
    \includegraphics[width=0.5\linewidth]{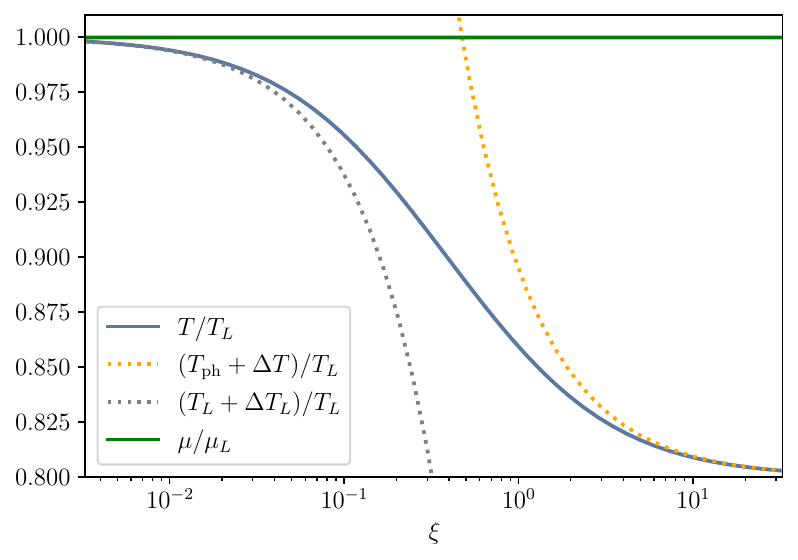}
    \caption{The temperature of the finite reservoir as the electron-phonon interaction is increased. Solid blue line shows the full numerical solution, while the orange dotted line and grey dotted line shows the approximations in Eqs.~\eqref{app:delta_T} and~\eqref{app:delta_T_L} respectively. The green solid line shows the chemical potential of the finite reservoir in the steady state. The parameters used are $T_L= T_R = 10\Gamma$, $V_b = 20\Gamma$, $V_g = 0$, $T_\mathrm{ph} = 8\Gamma$, and $\sigma = 1000\Gamma/T_L$.}
    \label{app:lin_res_T}
\end{figure}

We are now interested in the limit where the electron-phonon interaction is strong, i.e., when $\xi\gg 1$. In this regime, we expect the electron temperature to be close to the phonon temperature, $T = T_\mathrm{ph} + \Delta T$, with $\Delta T/T_\mathrm{ph} \ll 1$. If we further assume that the coupling to the infinite reservoir is much stronger than the coupling to the QD, $\zeta\gg 1$, we additionally have that $\mu = \mu_L + \Delta \mu$, with $\Delta \mu \ll 1$ so that the chemical potential of the finite reservoir is close to the chemical potential of the infinite reservoir. Analogously to Sec.~\ref{app:linres_big_sigma}, we can linearise Eqs.~\eqref{MuDot} and~\eqref{TDot} in $\Delta T$ and $\Delta \mu$, and solve for the shifts in temperature and chemical potential. Expanding to first order in $1/\xi$ yields
\begin{align}
    \Delta\mu &\approx -\frac{V_b}{\sigma}G\\
    \Delta T &\approx \frac{\pi^2T_{\mathrm{ph}}}{6n\xi}\left(1-\frac{T_{\mathrm{ph}}^2}{T_L^2}\right),\label{app:delta_T}
\end{align}
which is nothing but Eq.~\eqref{deltas} from the main text. 

Fig.~\ref{app:lin_res_T} shows the steady state temperature (solid blue line) as the electron-phonon interaction $\Sigma$ is increased. We also show the approximations in Eqs.~\eqref{app:delta_T} (orange dotted line) and~\eqref{app:delta_T_L} (gray dotted line). As expected, we see that when $\xi\ll 1$, the approximation in Eq.~\eqref{app:delta_T_L} agrees with the numerical result to an excellent degree. On the other hand, when $\xi\gg 1$, the approximation in Eq.~\eqref{app:delta_T} becomes accurate. Furthermore, we see that as $\xi\rightarrow\infty$, the electron temperature approaches $T_\mathrm{ph}$. Finally, we also show the steady state chemical potential of the finite reservoir (green solid line). We see that, as expected from the discussion above, $\mu$ is largely insensitive to $\xi$.

\section{Quantum Fisher information for a classical rate equation}\label{app:qfi}
In the following, we derive the ultimate sensitivity limit for any system $S$ described by the states $i=0,1,\dots,d-1$, $d$ being a positive integer, where a rate equation describes the transitions between states. The rate of transitions $i\rightarrow j$ is described by a rate matrix $\mathcal W_{ij}$, so the probability of the system being in state $i$ evolves in time as: 
\begin{align}
\label{app:rate_eq}
    \frac{\mathrm{d}p_i}{\mathrm{d}t} = \sum_{j}\mathcal{W}_{ij} p_j.
\end{align}
The transitions described in Eq.~\eqref{app:rate_eq} can, for example, be caused by the exchange of excitations with a thermal bath. We are interested in the ultimate precision limit for the estimation of a parameter $\theta$ that can be achieved by jointly measuring the global system (i.e., system and environment) during a time $\tau$. We assume that the system and environment are in the steady state. We use the framework developed in Ref.~\cite{Gammelmark2014} to show this. The first step is to map the classical rate equation into a Gorini--Kossakowski--Sudarshan--Lindblad (GKSL) master equation~\cite{Gorini1976, Lindblad1976}. The GKSL master equation takes the form:
\begin{align}
\label{eq:GKSL_master_eq}
    \frac{\mathrm{d}\hat{\rho}_S}{\mathrm{d}t} = \mathcal{L}[\hat{\rho}_S] 
    = -i[\hat{H}_S,\hat{\rho}_S] 
      + \sum_{k=1}^r \mathcal{D}[\hat{L}_k]\hat{\rho}_S.
\end{align}
Here, $\hat\rho_S$ is a diagonal matrix with entries $(\hat\rho_S)_{ii} = p_i$,
$\mathcal{L}$ is a superoperator called the Liouvillian, and $\hat{H}_S$ is the Hamiltonian of $S$. The first term on the right‑hand side of Equation~\eqref{eq:GKSL_master_eq} describes the unitary time evolution of $S$, while the dissipator 
\begin{align}
\label{eq:dissipator}
    \mathcal{D}[\hat{L}_k]\hat{\rho}_S 
    = \hat{L}_k\hat{\rho}_S\hat{L}_k^\dagger 
      - \frac{1}{2}\{\hat{L}_k^\dagger \hat{L}_k, \hat{\rho}_S\}
\end{align}
encodes the exchange of excitations between $S$ and the environment. Here $\{\hat A,\hat B\} = \hat A \hat B+\hat B\hat A$ denotes the anticommutator. Each jump operator $\hat{L}_k$ describes the exchange of excitations between the system and environment through the dissipation channel $k$. We can map the classical rate equation used in the main text into a GKSL master equation in the following way:
\begin{align}
\label{eq:master_equation_mapping}
    \mathcal W_{nj}\leftrightarrow \hat{L}_{k} = \sqrt{\mathcal W_{nj}}\dyad{n}{j}.
\end{align}

We are now interested in finding the maximum information we can extract from a joint measurement on $S$ and the environment state after a time $\tau$. To this end, we use the framework developed in Ref.~\cite{Gammelmark2014}, where it was shown that, in the long time limit, the quantum Fisher information scales linear in time, and is given by:
\begin{align}
\label{eq:molmer_Fisher_rate}
    \mathcal{F}_{T_\mathrm{ph}} = 4\tau\left[\sum_k \expval{\big(\partial_{T_\mathrm{ph}} \hat{L}_k\big)^\dagger \big(\partial_{T_\mathrm{ph}} \hat{L}_k\big)} - \Tr\left(\mathcal{L}_L \mathcal{L}^+\mathcal{L}_R [\hat{\rho}_S]  \right) - \Tr\left(\mathcal{L}_R \mathcal{L}^+\mathcal{L}_L [\hat{\rho}_S]  \right)  \right].
\end{align}
Here, $\mathcal{L}^+$ is the Drazin pseudo-inverse, and the superoperators $\mathcal{L}_{L/R}$ are defined as 
\begin{align}
    \mathcal{L}_L[\hat{\rho}] &= -i \big(\partial_{T_\mathrm{ph}}\hat{H}_\mathrm{eff}\big)\hat{\rho} + \sum_k\big(\partial_{T_\mathrm{ph}} \hat{L}_k\big)\hat{\rho}\hat L_k^\dagger,\\
    \mathcal{L}_R[\hat{\rho}] &= i\hat{\rho}\big(\partial_{T_\mathrm{ph}}\hat{H}_\mathrm{eff}^\dagger\big) + \sum_k \hat{L}_k \hat{\rho} \big(\partial_{T_\mathrm{ph}} \hat{L}_k^\dagger\big),
\end{align}
and we have introduced the (non-Hermitian) effective Hamiltonian $\hat{H}_\mathrm{eff} = \hat{H} - i/2 \sum_k\hat{L}_k^\dagger \hat{L}_k$. By inserting the form of the jump operators in Eq.~\eqref{eq:master_equation_mapping} and performing some tedious, but straightforward algebra, we get that for any system described by a classical rate equation, the QFI is given by:
\begin{align}
\label{app:QFI_rate_equation}
    \mathcal{F}_{T_\mathrm{ph}}= \tau\sum_{i,j} \frac{(\partial_{T_\mathrm{ph}} \mathcal W_{ij})^2}{\mathcal W_{ij}}p_j,
\end{align}
which is exactly the same as the Fisher information obtained by continuously monitoring the transitions and of $S$~\cite{Smiga2023}. 

\section{Calculation of $F_I$}\label{app:F_I}
In this Section, we derive Eq.~\eqref{ApproxF_I} from the main text. The starting point will be Eq.~\eqref{app:QFI_rate_equation}. Inserting the form of $\mathcal{W}$ from the one used in the main text, and using that the rates from the QD into the right reservoir are independent of $T_\mathrm{ph}$, we get that the QFI is given by:
\begin{align}
\label{app:QFI_molmer}
    F_I(T_{\mathrm{ph}}) =  \left[(\partial_{T_{\mathrm{ph}}}{\Gamma_L^{\mathrm{in}}})^2  \frac{p_0^{\mathrm{ss}}}{\Gamma_L^{\mathrm{in}}} + (\partial_{T_{\mathrm{ph}}}\Gamma_L^{\mathrm{out}})^2  \frac{p_1^{\mathrm{ss}}}{\Gamma_L^{\mathrm{out}}}\right] \tau.
\end{align}
As described in Secs.~\ref{app:linres_big_sigma} and~\ref{app:linres_big_Sigma}, in the limit when $\zeta\gg 1$ the chemical potential is insensitive to the phonon temperature, i.e., $\partial_{T_\mathrm{ph}}\mu \approx 0$ in this regime. Using the chain rule, we can immediately see that this means that we can write the derivatives in Eq.~\eqref{app:QFI_molmer} as
\begin{align}
    \frac{\partial \Gamma_L^\mathrm{in}}{\partial T_\mathrm{ph}} = \Gamma \frac{\partial f}{\partial T} \frac{\partial T}{\partial T_\mathrm{ph}},
\end{align}
and similarly for the derivative of $\Gamma_L^\mathrm{out}$. We then define the quantities 
\begin{align}
    \chi_e &= \left(\frac{\partial f}{\partial T} \right)^2 = \frac{(\epsilon -\mu)^2 }{16T^4\cosh^4(\frac{\epsilon - \mu}{2T})},\\
    \chi_\mathrm{ph} &= \left(\frac{\partial T}{\partial T_\mathrm{ph}} \right)^2,
\end{align}
as in the main text. To make further progress, we consider the two regimes of when $\xi\gg 1$ and $\xi \ll 1$, and use the approximate expressions for $T$ introduced in Eqs.~\eqref{app:delta_T} and~\eqref{app:delta_T_L} to write $\chi_\mathrm{ph}$ as:
\begin{align}
\label{app:chi_ph}
    \chi_\mathrm{ph} \approx \begin{cases}
        \left[1 +\frac{\pi^2}{6n\xi}\left((1-n)+\frac{T_{\mathrm{ph}}^2}{T_L^2}(n-3)\right)\right]^2, \quad \xi\gg 1\\
        \left( \frac{3n\xi T_L}{\pi^2 T_\mathrm{ph}} \right)^2, \qquad\qquad\qquad\qquad\qquad\qquad\!\xi\ll1
    \end{cases}
\end{align}
This allows us to approximate $F_I$ in both the regimes of weak and strong electron-phonon interaction. The crucial assumption in deriving both expressions in Eq.~\eqref{app:chi_ph} is that $\zeta\gg 1$, so that the chemical potential is insensitive to the phonon temperature. In the main text, we focus on the regime $\xi\gg1$, as this is where we expect good sensitivity.

\section{Calculation of $F_{II}$}\label{appx:F_II}
In this section, we derive the expression for $F_{II}$ used in the main text. We consider a measurement strategy where we measure the total charge transferred from left to right in a time $\tau$. This can be accomplished, e.g., by making a two-point measurement on the total charge in the right reservoir $\hat{N}(\tau) = \hat{N}_R(\tau) - \hat{N}_R(0)$. The Fisher information of such observables was studied in Ref.~\cite{Khandelwal2025}, where they found that, in the long time limit, the Fisher information is given by:
\begin{align}
\label{app:F_II}
    F_{II} = \frac{\tau \left(\partial_{T_\mathrm{ph}} \expval{I_\mathrm{QD}}\right)^2}{\langle\!\langle I_\mathrm{QD}\rangle\!\rangle^2}.
\end{align}
By neglecting fluctuations in the finite reservoir temperature and chemical potential, we can calculate the diffusion coefficient as (see sec. IV D in Ref.~\cite{Landi2024} for details on how to compute this)
\begin{align}
     \langle\!\langle {I_{\mathrm{QD}}}^2 \rangle\!\rangle = \Gamma\left[  \frac{f\bar{f}+f_R\bar{f}_R}{2} + \frac{(f_R - f)^2}{4}\right].
\end{align}
Furthermore, since $\expval{I_\mathrm{QD}} = \Gamma(f-f_R)/2$, we have that the derivative of the average current $\partial_{T_\mathrm{ph}}\expval{I_\mathrm{QD}} = \Gamma \partial_{T_\mathrm{ph}}f$. Using the same techniques as in sec.~\ref{app:F_I}, we get that the Fisher information is given by
\begin{align}
    F_{II}(T_{\mathrm{ph}}) \approx \chi_e \chi_{\mathrm{ph}} \frac{\Gamma \tau}{(f+f_R)(\bar{f} + \bar{f}_R)},
\end{align}
as in the main text.

\end{document}